\documentclass[12pt]{article}
\usepackage{amsfonts}
\usepackage{latexsym}
\usepackage{amsmath}
\usepackage{amssymb}
\usepackage{amssymb}

\newcommand{\newsection}{    
\setcounter{equation}{0}\section}
\def\appendix#1{\addtocounter{section}{1}\setcounter{equation}{0}
\renewcommand{\thesection}{\Alph{section}}
\section*{Appendix \thesection\protect\indent \parbox[t]{11.15cm}{#1}}
\addcontentsline{toc}{section}{Appendix \thesection\ \ \ #1}}

\newcommand{\be}{\begin{eqnarray}}
\newcommand{\ee}{\end{eqnarray}}
\newcommand{\bea}{\begin{eqnarray}}
\newcommand{\eea}{\end{eqnarray}}
\newcommand{\ba}{\begin{array}}
\newcommand{\ea}{\end{array}}
\newcommand{\nn}{\nonumber \\}

\newenvironment{frcseries}{\fontfamily{frc}\selectfont}{}

\def\cL{{\cal L}}

\def\bo{{\bar{1}}}
\def\bt{{\bar{2}}}
\def\lp{{\lambda^1_+}}
\def\lm{{\lambda^1_-}}
\def\lpb{{\lambda^{\bar{1}}_+}}
\def\lmb{{\lambda^{\bar{1}}_-}}
\def\hN{{\hat{N}}}

\begin{document}
\begin{titlepage}
\begin{center}
\vspace*{-1.0cm}
\hfill DMUS--MP--13/15 \\

\vspace{3.0cm} {\Large \bf Supersymmetric AdS Black Rings} \\[.2cm]

\vskip 2cm
J. Grover$^1$, J. B.  Gutowski$^2$ and W. A. Sabra$^3$
\\
\vskip .6cm

\begin{small}
$^1$\textit{Physics Department, University of Aveiro and I3N, \\
Campus de Santiago,
3810-183 Aveiro, Portugal \\
E-mail: jai@ua.pt}
\end{small}\\*[.6cm]

\begin{small}
$^2$\textit{Department of Mathematics,
University of Surrey \\
Guildford, GU2 7XH, UK \\
E-mail: j.gutowski@surrey.ac.uk}
\end{small}\\*[.6cm]

\begin{small}
$^3$\textit{Centre for Advanced Mathematical Sciences and
Physics Department, \\
American University of Beirut, Lebanon \\
E-mail: ws00@aub.edu.lb}
\end{small}\\*[.6cm]

\end{center}

\vskip 3.5 cm
\begin{abstract}

It has been proven in arXiv:1303.0853 that all regular supersymmetric near-horizon geometries in
minimal five-dimensional gauged supergravity admit automatic supersymmetry
enhancement. Using this result, the integrability conditions associated with
the existence of the additional supersymmetry are analysed, and the
near-horizon geometries are determined explicitly.
We show that they all correspond to previously constructed examples.
Hence, there are no supersymmetric black ring solutions in minimal five-dimensional gauged supergravity.

\end{abstract}

\end{titlepage}



\section{Introduction}

The study of supersymmetric gravitational solutions has many applications in string theory, in establishing the uniqueness or non-uniqueness of solutions, and in the discovery of new exotic solutions in several supergravity theories. An example of exotic solutions are the black rings of five dimensional ungauged supergravity theories \cite{robharv}.  In our present work we will be mainly concerned with minimal $N=2$,  $D=5$ gauged supergravity. In this theory, supersymmetric solutions can in principle preserve 1/4, 1/2, 3/4 or the maximal proportion of supersymmetry.

Examples of regular asymptotically $AdS_5$ black holes preserving 1/4 supersymmetry are found in \cite{adsbh,gutreal04}.
These solutions were later generalized in \cite{chong1, chong2, kun006}. 1/4-supersymmetric string solutions have also been constructed in \cite{chamsabra, klemmsabra}. In \cite{gauntgut03} a systematic classification of all 1/4-supersymmetric solutions of minimal gauged $N=2$, $D=5$ supergravity was performed, this was later extended to more general theories with Abelian vector multiplets \cite{gutsab03}. Examples of 1/2-supersymmetric solutions,  corresponding to domain walls and black holes without regular horizons, were constructed in \cite{halfsup}. In \cite{preon1}, it was shown that all 3/4-supersymmetric solutions are locally $AdS_5$. However, globally one can have discrete quotients of  $AdS_5$ as 3/4-supersymmetric configurations \cite{preon2}. The unique maximally supersymmetric solution preserving all supersymmetries is $AdS_5$. Systematic classifications of 1/2-supersymmetric solutions were investigated in \cite{halfsys}, \cite{halfnull} using spinorial geometry techniques first implemented in the study of higher dimensional solutions in \cite{spin1, spin2, spin3}. Using these methods, a restricted class of half-supersymmetric near-horizon geometries was analysed in \cite{enhanced2}.

An interesting phenomenon in supersymmetric black hole physics is the enhancement of supersymmetry for near-horizon geometries. This has been established for many supersymmetric black hole and brane solutions.
It is believed that it may hold for all supersymmetric black hole solutions, at least in theories for
which there are no higher order curvature corrections.
Clearly supersymmetry enhancement puts more restrictions on the topology and geometry of horizons and thus may lead to an explicit determination of the space-time metric. Recently, it was demonstrated in \cite{enhance} that under some smoothness assumptions, supersymmetric near-horizon black hole geometries of minimal 5-dimensional gauged supergravity preserve at least half of the supersymmetry. Near-horizon geometries with more than half of the supersymmetries preserved are locally isomorphic to $AdS_5$, with vanishing Maxwell field strengths.

The analysis presented in this paper is more general than that carried out in \cite{enhanced2}.
This is because in \cite{enhanced2} it was assumed that the
event horizon corresponds to a null hypersurface for {\it all} the Killing vectors obtained from
Killing spinor bilinears. Here, we shall not make this assumption. We assume that
the event horizon is a null hypersurface associated with one of the Killing vectors obtained
from one of the Killing spinors, but is not necessarily a null hypersurface with respect
to all of the Killing vectors obtained from all of the Killing spinors. In particular, we shall be interested
in the case for which one of the Killing spinors exists not only for the near-horizon geometry, but can
be extended to give a Killing spinor for the bulk solution, whereas the extra Killing spinor of the near-horizon solution need not
extend to a bulk Killing spinor. We assume that in the near-horizon solution,
the event horizon corresponds to a null hypersurface associated with the Killing vector obtained
from the Killing spinor
which extends out to the bulk. However, we shall not assume, as was done previously, that
the extra Killing spinor produces a Killing vector which also becomes null at the event horizon.

In this paper, using the results of \cite{enhance} and employing the integrability conditions associated with the extra supersymmetry, the near-horizon geometries with enhanced supersymmetry are determined explicitly. These geometries all correspond to previously known examples, and thus it can be established that there are no supersymmetric black ring solutions in minimal five-dimensional gauged supergravity.

We organise our work as follows. In section two, we summarize the results of \cite{adsbh} where conditions for 1/4 supersymmetric near-horizon geometries in minimal gauged $D=5$ supergravity were obtained. In section three, using the results of \cite{enhance}, we analyse the necessary conditions for the enhancement of supersymmetry at the horizon. These conditions are sufficient to fix the near-horizon geometry completely,
and they all correspond to solutions which are already known. We make use of the the Killing spinor equations, which are analaysed using spinorial geometry techniques presented in Appendix A, and the associated integrability conditions are presented in Appendix B.

\newsection{The $1/4$-Supersymmetric Near-Horizon Solution}

In this section, we briefly summarize the conditions on the event horizon of
a regular black hole in minimal gauged $D=5$ supergravity which
are imposed by requiring that the solution should preserve $1/4$ of the supersymmetry \cite{adsbh}.
This is the minimal amount of supersymmetry which supersymmetric solutions
of this theory can preserve.

The near-horizon metric is given by
\be
ds^2 = -r^2 \Delta^2 du^2 +2 du dr + 2r h du + ds_{\cal{S}}^2 \ .
\ee
The horizon is at $r=0$, $ds_{\cal{S}}^2$ is the metric on the (compact) and
regular spatial cross-section of the horizon, which does not depend
on $u$ or $r$; $h$ is a 1-form on
${\cal{S}}$ which does not depend on $u$ or $r$; and $\Delta$ is a function
which does not depend on $u$ or $r$.

In the notation of \cite{adsbh}, a (real) basis for ${\cal{S}}$ is given by $Z^1, Z^2, Z^3$.
The covariant derivatives of the basis elements are fixed by
\begin{equation}
\label{dzeqn}
\nabla_A Z^i_B = -{\Delta \over 2}(\star Z^i)_{AB} + \gamma_{AB}\big(N \cdot Z^i + \frac{1}{\ell} \delta_{1i} \big) - Z^i_{A}N_{B} - \frac{Z^i_{A}Z^1_{B}}{\ell} + 2 \sqrt{3} \ell^{-1} \epsilon_{1ij} a_A Z^j_B,
\end{equation}
where $\star$ denotes the Hodge dual on ${\cal{S}}$ with positive orientation defined by $Z^1 \wedge Z^2 \wedge Z^3$, and
\be
N = h + {2 \over \ell} Z^1 \ ,
\ee
and $a$ is a $u, r$-independent 1-form on ${\cal{S}}$ satisfying
\be
\label{daeqn}
d a = -{\sqrt{3} \over 2} \star N \ .
\ee
The 2-form flux is given by
\be
F = -{\sqrt{3} \over 2} \big( du \wedge d (r \Delta) + \star N \big)
\ee
and the Bianchi identity implies that
\be
d \star N =0 \ ,
\ee
and $N$ also satisfies
\be
\label{dneq}
dN = 2 \ell^{-1} \Delta \star_3 Z^1 + 2 \ell^{-1} N \wedge Z^1+ \Delta \star N + \star d \Delta \ .
\ee
These conditions constrain the Ricci tensor of ${\cal{S}}$ to be
\begin{equation}
\label{ricsc}
R_{AB} = \gamma_{AB}\left(\frac{\Delta^2}{2} + N^2 + \frac{2}{\ell}(N\cdot Z^1) - \frac{2}{\ell^2}\right) - \nabla_{(A}N_{B)} - \frac{4}{\ell}Z^1_{(A}N_{B)} - N_AN_B \ .
\end{equation}
For convenience, we shall set $Z=Z^1$.

\subsection{Solutions with $N=0$}

Before proceeding to examine necessary conditions for enhanced supersymmetry,
it is useful to consider a special class of solutions for which $N=0$ in some open set. These have been
analysed previously in \cite{adsbh}; we again summarize the analysis here for convenience. If $N=0$, then
({\ref{ricsc}}) implies that
\be
R_{AB}= ({1 \over 2} \Delta^2-2 \ell^{-2}) \gamma_{AB}
\ee
and hence the curvature Bianchi identity implies that $d \Delta^2=0$,
and together with ({\ref{dneq}}) this implies that $\Delta=0$. Then, ({\ref{dzeqn}}) implies that one
can introduce local coordinates $x, y, z$ on ${\cal{S}}$ such that
\be
Z =dx
\ee
and
\be
ds_{\cal{S}}^2 = dx^2 + e^{2x \over \ell} (dy^2+dz^2)
\ee
so that the five-dimensional spacetime metric is
\be
ds^2 = 2 du dr -{4 \over \ell} r dx du +  dx^2 + e^{2x \over \ell} (dy^2+dz^2) \ .
\ee
Hence, the spacetime is locally isometric to $AdS_5$. We have remarked previously that there exist discrete
quotients of $AdS_5$ which are 3/4-supersymmetric. There may also be quotients of $AdS_5$ which preserve
less supersymmetry. In the following analysis, we shall be concerned with geometries other than
$AdS_5$ or its quotients.

\newsection{Analysis of Supersymmetry}

To proceed, we shall consider some necessary conditions imposed by
the enhancement of supersymmetry, which follows as a consequence of
the analysis in \cite{enhance}.  We will demonstrate that these conditions
can be used to completely determine the near-horizon geometries.
The analysis of the Killing spinor equations
is carried out in Appendix A, using spinorial geometry techniques,
and their associated integrability conditions are investigated in Appendix B.
We exclude the solutions which are (locally) $AdS_5$ with $F=0$.

We begin by considering the conditions ({\ref{ex1}})-({\ref{ex7}}).
There are two possibilities; either $\Delta$ is not identically zero,
or $\Delta=0$ everywhere on ${\cal{S}}$. We shall consider these two cases separately.
When $\Delta$ is not identically zero, we restrict to a 
patch on which, without loss of generality, $\Delta >0$.
If $\Delta$ is analytic on the horizon then it was shown in \cite{KLR}
that either $\Delta=0$ everywhere on the horizon, or $\Delta>0$
everywhere on the horizon. However, in what follows it suffices to
assume that $\Delta$ is smooth.

\subsection{Solutions with $\Delta \neq 0$}

We first consider the case for which $\Delta>0$ on some patch in ${\cal{S}}$.
Let $P$ be some point in ${\cal{S}}$; there are two sub-cases to consider.
In the first,  $N \neq 0$ at $P$.
Then the condition ({\ref{ex1}}) is non-trivial.
So, for solutions with supersymmetry enhancement,
it is clear that the LHS of each of the conditions ({\ref{ex2}})-({\ref{ex7}}) must be proportional
to the LHS of the condition ({\ref{ex1}}). Hence we obtain six conditions which are necessary
and sufficient for the integrability conditions to have a non-trivial solution
for $\lm, \mu$, other than that corresponding to $AdS_5$. Each of these six conditions
come from requiring that the determinant of six $2 \times 2$ matrices vanish, the first row
of the matrices coming from the the LHS of ({\ref{ex1}}) and the second rows
coming from the LHS of each of  ({\ref{ex2}})-({\ref{ex7}}). A detailed analysis of these conditions shows that they
are equivalent to imposing

\bea
\label{xcon1}
\nabla_{(i} N_{j)} = {2 \ell^{-1} \over \Delta^2+N^2} \bigg( \Delta(N_i \star_3 (Z \wedge N)_j
+ N_j \star_3 (Z \wedge N)_i)
\nn
+2 (N.Z) N_i N_j -N^2 Z_i N_j -N^2 Z_j N_i \bigg)
\eea
and
\bea
\label{xcon2}
d \Delta = {4 \Delta \ell^{-1} \over \Delta^2+N^2} \bigg( \Delta \star_3 (Z \wedge N) - N^2 Z + (N.Z) N \bigg) \ .
\eea

In the second case, $N=0$ at $P$.
Then in order for there to be enhanced supersymmetry, ({\ref{ex1}}) implies that $\lambda^1_- =0$,
$\mu \neq 0$ at $P$; ({\ref{ex2}}) and ({\ref{ex3}}) imply that $d \Delta=0$ at $P$, and
({\ref{ex4}})-({\ref{ex7}}) imply that $\nabla_{(i} N_{j)}=0$ at $P$. Hence, ({\ref{xcon1}}) and ({\ref{xcon2}})
hold at points for which $N=0$ as well.

In particular,
the conditions ({\ref{xcon1}}) and ({\ref{xcon2}}) imply that
\be
\label{scaldelt1}
\Delta^{-{3 \over 2}} N^2 + \Delta^{1 \over 2}  = \nu_1
\ee
and
\be
\label{scaldelt2}
\Delta^{-1} \bigg( N.Z - {\ell \over 2} (N^2+ \Delta^2) \bigg) = \nu_2
\ee
for constants $\nu_1, \nu_2$.

We shall first consider the case for which $N^2-(N.Z)^2 \neq 0$ in some patch.
Then $N$ and $Z$ are linearly independent in this patch; the case for which
$N$ is proportional to $Z$ will be considered later.

It is convenient to define
\be
{\hat{N}} = - \Delta Z + \star (N \wedge Z)
\ee
and note that $N$, ${\hat{N}}$ are linearly independent.
Consider the $1/4$ supersymmetry conditions together with the enhancement conditions
({\ref{xcon1}}) and ({\ref{xcon2}}). It is straightforward to see
 that ${\hat{N}}= \star dZ$, so ${\hat{N}}$ is co-closed.
Furthermore we find the conditions
\be
\nabla_{(i} (\Delta^{-1} N)_{j)}=  \nabla_{(i} (\Delta^{-1} {\hat{N}})_{j)}= 0
\ee
so $\Delta^{-1} N$ and $\Delta^{-1} \hN$ are Killing vectors, in fact they also commute with each other.
Hence, we introduce
co-ordinates $y^1, y^2$ on the horizon such that
\be
V = \Delta^{-1} N = {\partial \over \partial y^1}, \qquad W = \Delta^{-1} \hN = {\partial
\over \partial y^2}
\ee
and note that
\be
\cL_V \Delta = \cL_W \Delta =0 \ .
\ee
In addition, as ({\ref{xcon2}}) implies that $d \Delta \neq 0$, one can take  $y^1, y^2, \Delta$
as local co-ordinates on ${\cal{S}}$. One then obtains the horizon metric explicitly as:
\bea
ds_{\cal{S}}^2 &=& {1 \over 16} \nu_1 \ell^2 \Delta^{-2} \big(\nu_1 -(1+\nu_2^2)\Delta^{1 \over 2}
- \nu_1 \nu_2 \ell \Delta - {1 \over 4} \nu_1^2 \ell^2 \Delta^{3 \over 2} \big)^{-1} (d \Delta)^2
\nn
&+& (\nu_1 \Delta^{-{1 \over 2}} -1) (dy^1)^2 -2 (\nu_2 +{1 \over 2} \nu_1 \ell \Delta^{1 \over 2})
dy^1 dy^2
\nn
&+& \big(\nu_1 \Delta^{-{1 \over 2}} - \nu_2^2 - \nu_1 \nu_2 \ell \Delta^{1 \over 2} -{1 \over 4} \nu_1^2 \ell^2
\Delta \big) (dy^2)^2
\eea
with
\be
Z =  (\nu_2 + {1 \over 2} \nu_1 \ell \Delta^{1 \over 2}) dy^1 - dy^2 -{1 \over 4} \ell \Delta^{-1} d \Delta \ ,
\ee
$\nu_1$ and $\nu_2$ are the real constants which appear in the conditions ({\ref{scaldelt1}}),
({\ref{scaldelt2}}).

It is straightforward to show that this near-horizon data is identical to the near-horizon
data for the Chong et al. solution presented in \cite{KLR, chong1, chong2}, in the case when
$\Delta \neq 0$. To establish the correspondence, set
\bea
\Delta&=& {\Delta_0 \over \Gamma^2}
\nn
\nu_1 &=& 4 \ell^{-2} C^{-2} \Delta_0^{1 \over 2} + C^2 \alpha_0^2 \Delta_0^{-{3 \over 2}}
\nn
\nu_2 &=& -{1 \over 2} C^2 \ell \alpha_0 \Delta_0^{-1}
\nn
y^1 &=& {1 \over 4 \Delta_0^2 + C^4 \ell^2 \alpha_0^2} \big( \ell^2 C^4 \Delta_0 \alpha_0 x^1
+ 4 \Delta_0^2 x^2 \big)
\nn
y^2 &=&  {1 \over 4 \Delta_0^2 + C^4 \ell^2 \alpha_0^2} \big( 2 C^2 \ell \Delta_0^2 x^1 - 2 \ell C^2 \alpha_0
\Delta_0 x^2 \big) \ .
\eea

Next, consider the special case for which $N^2-(N.Z)^2 = 0$. Then $N=gZ$ for some function $g$.
It is straightforward to see that ({\ref{xcon2}}) implies that $\Delta$ is constant.
On substituting these conditions into ({\ref{dzeqn}}) and ({\ref{dneq}}) one finds that
\be
N = - \ell^{-1} Z
\ee
and furthermore, an examination of the conditions and ({\ref{dzeqn}}), ({\ref{daeqn}}) implies that
one can without loss of generality choose $Z^1, Z^2, Z^3$ to satisfy
\bea
d Z^1 &=& - \Delta Z^2 \wedge Z^3
\nn
dZ^2 &=& \Delta (1-3 \ell^{-2} \Delta^{-2}) Z^1 \wedge Z^3
\nn
dZ^3 &=& - \Delta (1-3 \ell^{-2} \Delta^{-2}) Z^1 \wedge Z^2 \ ,
\eea
and hence the metric on ${\cal{S}}$ is either that of the homogeneous metric on the Nil-manifold
(when $\Delta = \sqrt{3} \ell^{-1}$),  the homogeneous metric on the $SL(2,R)$ group manifold (when
$0<\Delta<\sqrt{3}\ell^{-1}$), or the homogeneous metric on the $SU(2)$ group manifold
(when $\Delta > \sqrt{3} \ell^{-1}$).
We remark that these metrics (and the corresponding spacetime geometries) have been previously derived in
\cite{adsbh}.

\subsection{Solutions with $\Delta =0$}

Next we consider the special case of solutions with $\Delta=0$.
For solutions with $\Delta=0$ on ${\cal{S}}$, consider some point
$P \in {\cal{S}}$. There are again two sub-cases. In the first, $N \neq 0$ at $P$,
and it is straightforward to show that the necessary and sufficient conditions for enhanced supersymmetry
is ({\ref{xcon1}}) with $\Delta=0$. In the second sub-case, for which $N=0$ at $P$,
the conditions ({\ref{ex1}})-({\ref{ex3}}) have no content, and ({\ref{ex4}})-({\ref{ex7}})
imply that $\nabla_{(i} N_{j)}=0$ at $P$.

So, if $\Delta=0$, then the necessary and sufficient conditions for enhanced supersymmetry are:
\begin{equation}
\label{xcon3}
\nabla_{(i} N_{j)} = \begin{cases} {2 \ell^{-1} \over N^2} \bigg(2 (N.Z) N_i N_j -N^2 Z_i N_j -N^2 Z_j N_i \bigg) \qquad &{\rm if} \ \ N^2 \neq 0 \cr 0 &{\rm if} \ \ N=0  \ . \end{cases}
\end{equation}

As we have assumed that $N$ is not identically zero, consider some patch in which
$N^2 \neq 0$. Then ({\ref{xcon3}}), together with the conditions in 
Section 2, implies that
\begin{equation}
d N^2 = {6 \over \ell} \bigg( (N.Z) N - N^2 Z \bigg)
\end{equation}
and hence
\begin{equation}
\nabla^i \nabla_i N^2 +{6 \over \ell} Z^i \nabla_i N^2
+{12 \over \ell}((N.Z)+\ell^{-1}) N^2 =0 \ .
\end{equation}
So, if $N^2$ is analytic on ${\cal{S}}$, then by the results of \cite{KLR}
it follows that $N^2$ must be nonzero everywhere on ${\cal{S}}$, because if
$N^2$ were to vanish at some point in ${\cal{S}}$, then $N$ must vanish identically.
In what follows, it suffices to assume that $N^2 \neq 0$ in some patch on ${\cal{S}}$.

In addition, ({\ref{xcon3}}) also implies that
\be
\label{scalconst1}
(N^2)^{-{2 \over 3}} (N.Z) - {\ell \over 2} (N^2)^{1 \over 3} = \nu
\ee
where $\nu$ is constant.

Again, we first consider solutions for which $N^2-(N.Z)^2 \neq 0$,
so that $N$ and $Z$ are linearly independent; the case for which
$N$ is proportional to $Z$ will be considered later.
It is convenient to define
\be
{\hat{N}} =  \star (N \wedge Z)
\ee
and note that $N$, ${\hat{N}}$ are linearly independent; and ${\hat{N}}$ is co-closed.
Then it is straightforward to show that the conditions imposed by $1/4$-supersymmetry,
together with the supersymmetry enhancement condition ({\ref{xcon3}}) imply that
\be
\nabla_{(i} ((N^2)^{-{2 \over 3}} N)_{j)}=\nabla_{(i} ((N^2)^{-{2 \over 3}} \hN)_{j)} =0
\ee
so $(N^2)^{-{2 \over 3}} N$, $(N^2)^{-{2 \over 3}} {\hat{N}}$ are Killing; and furthermore,
the Killing vectors commute.

Hence, we introduce
co-ordinates $y^1, y^2$ on the horizon such that
\be
V = (N^2)^{-{2 \over 3}} N = {\partial \over \partial y^1}, \qquad W = (N^2)^{-{2 \over 3}}  \hN = {\partial
\over \partial y^2} \ .
\ee
In addition, ({\ref{xcon3}}) implies that
\be
\cL_V N^2 = \cL_W N^2 =0
\ee
and $d N^2 \neq 0$. We will therefore also use $N^2$ as a co-ordinate on the horizon.

It is then straightforward to show that, in the co-ordinates $y^1, y^2, N^2$, the  metric
on the horizon is
\bea
ds_{\cal{S}}^2 &=& {1 \over 36} \ell^2 (N^2)^{-2} \big( 1-{1 \over 4} \ell^2 N^2 - \ell \nu (N^2)^{2 \over 3}
- \nu^2 (N^2)^{1 \over 3} \big)^{-1} (dN^2)^2
\nn
&+& (N^2)^{-{1 \over 3}} (dy^1)^2 + (N^2)^{-{1 \over 3}}
\big( 1-{1 \over 4} \ell^2 N^2 - \ell \nu (N^2)^{2 \over 3}
- \nu^2 (N^2)^{1 \over 3} \big) (dy^2)^2
\eea
with
\be
Z =(\nu + {1 \over 2} \ell (N^2)^{1 \over 3}) dy^1 - {1 \over 6} \ell (N^2)^{-1} d N^2 \ ,
\ee
$\nu$ is the real constant which appears in the condition ({\ref{scalconst1}}).

It is straightforward to show that this near-horizon data is identical to that corresponding to the solution presented in \cite{KLR}, in the case when
$\Delta_0=0$, and $\alpha_0 \neq 0$. To establish the correspondence, set
\bea
\Gamma &=& C^{2 \over 3} \alpha_0^{2 \over 3}  (N^2)^{-{1 \over 3}}
\nn
\nu &=& -{1 \over 2} C^{2 \over 3} \ell \alpha_0^{-{1 \over 3}}
\nn
y^1 &=& C^{4 \over 3} \alpha_0^{1 \over 3} x^1
\nn
y^2 &=& - 2 \ell^{-1} C^{-{2 \over 3}} \alpha_0^{1 \over 3} x^2 \ .
\eea

Finally, consider the special case when $N^2-(N.Z)^2 = 0$, so $N = g Z$ for some function $g$.
Note that ({\ref{xcon3}}) implies that $N$ is a Killing vector. This condition, together with the conditions imposed
by ({\ref{dzeqn}}), implies that $g$ is constrained to be constant, and on discarding the case
$g=0$ (which corresponds to $AdS_5$), one obtains
\be
N = - \ell^{-1} Z \ .
\ee
Then the conditions ({\ref{dzeqn}}) and ({\ref{daeqn}}) imply that one can choose a local co-ordinate
$x, y, z$ on ${\cal{S}}$ such that
\be
Z = dx, \qquad \qquad ds_{\cal{S}}^2 = dx^2  +{\ell^2 \over 3 y^2} (dy^2+dz^2)
\ee
so that $ds_{\cal{S}}^2$ is the metric on $R \times H^2$. This solution was also previously found in
\cite{adsbh}.

\newsection{Conclusions}

We have analysed the integrability conditions obtained from the Killing spinor equations
for near-horizon geometries in minimal five-dimensional gauged supergravity.
By making use of the fact that near-horizon solutions other than $AdS_5$ must be
exactly half-supersymmetric \cite{enhance}, we show that these integrability conditions
are  strong enough to completely determine the near-horizon geometries.
We demonstrate that all of the near-horizon solutions correspond to examples
which have already been obtained in \cite{adsbh} and \cite{KLR}.
It should be noted that although we have explicitly given the co-ordinate transformations required
in order to recover the 
black hole near-horizon geometries obtained from
\cite{KLR} for the sake of completeness,
this is not strictly necessary. This is because once one has obtained the two
commuting isometries $\Delta^{-1}N, \Delta^{-1} {\hat{N}}$
and $(N^2)^{-{2 \over 3}} N, (N^2)^{-{2 \over 3}} {\hat{N}}$ in sections 3.1 and 3.2 respectively,
the near-horizon geometries are then fixed using the results of \cite{KLR}.

As none of the near-horizon geometries contains a horizon section ${\cal{S}}$ which is
topologically $S^1 \times S^2$, this implies that there are no regular supersymmetric
asymptotically $AdS_5$ black rings in minimal five dimensional supergravity.
The analogous result for pseudo-supersymmetric asymptotically $dS_5$
black rings in minimal five dimensional supergravity has already been established in
\cite{desit}. In this case, all (pseudo)-supersymmetric solutions automatically preserve
at least half of the supersymmetry in contrast to the theory with a 
negative cosmological constant,
so the analysis was more straightforward. 

The status of supersymmetric asymptotically
$AdS_5$ or $dS_5$ black rings in non-minimal gauged supergravity, for example
coupled to some abelian vector multiplets, remains undetermined.
As we have proven that such solutions do not exist in the minimal theories,
they must lie in a sector of the theory which does
not admit a reduction to the minimal theory. 
It is known that there exist supersymmetric near-horizon
geometries with horizon section $S^1 \times S^2$ in 
gauged supergravity with a negative cosmological constant,
which cannot be reduced to a solution of the minimal theory,
\cite{nonminimal}.
However, it is not yet known if a full black ring solution
in this theory exists. Work on such solutions is in progress.

\vskip 0.5cm
\noindent{\bf Acknowledgements} \vskip 0.1cm
\noindent 
 J. Gutowski is supported by the STFC grant, ST/1004874/1.  J. Grover is supported by FCT-Portugal, under the grant SFRH/BPD/78142/2011, and the project \newline PTDC/FIS/116625/2010, as well as by the NRHEP295189  FP7-PEOPLE-2011-IRSES Grant. The authors would like to thank H. Kunduri, J. Lucietti and H. Reall for useful discussions.
\vskip 0.5cm

\setcounter{section}{0}

\appendix{The Killing Spinor Equations}

The Killing spinor equations adapted to a null basis have been
computed in Appendix B of \cite{halfnull}, using spinorial geometry techniques.
These were originally developed to classify solutions in ten and eleven dimensional
supergravity theories, \cite{spin1, spin2, spin3}, and can also be adapted to the classification of
supersymmetric black holes. We use the same conventions
for the Dirac Killing spinors, which are elements of the set of all complexified forms
on $R^2$, spanned by $1, e_1, e_2$ and $e_{12}=e_1 \wedge e_2$.
A generic Killing spinor is then written as
\be
\label{genspin}
\epsilon = \lambda^1_+ (1+e_1) +\lambda^1_- (1-e_1) +\lambda^{\bar{1}}_+ (e_{12}-e_2)
+ \lambda^{\bar{1}}_- (e_{12}+e_2) \ .
\ee

We work with a basis in which
the metric is:
\be
ds^2 = -2 {\bf{e}}^+ {\bf{e}}^- + ({\bf{e}}^1)^2 +2 {\bf{e}}^2 {\bf{e}}^{\bar{2}}
\ee
where ${\bf{e}}^+, {\bf{e}}^-, {\bf{e}}^1$ are real, and ${\bf{e}}^2$, ${\bf{e}}^{\bar{2}}$ are
complex conjugate. We shall investigate the supersymmetry of regular near-horizon geometries
using the conventions set out in \cite{adsbh}.
In particular, note that in order to rewrite the Killing spinor equations
computed in \cite{halfnull} in terms of the conventions
adopted in \cite{adsbh}, we make the following replacements:
\bea
\chi \rightarrow {2 \sqrt{3} \over \ell}, \qquad
\chi V_I X^I \rightarrow {1 \over \ell}, \qquad
H \rightarrow {2 \over \sqrt{3}} F, \qquad
\chi A \rightarrow {2 \over \sqrt{3} \ell} A
\eea
and we also re-label the basis indices as
\be
1 \rightarrow 2, \qquad \bo \rightarrow \bt, \qquad 2 \rightarrow 1
\ee
and make a sign change to the spin connection
\be
\omega_{\mu_1, \mu_2 \mu_3} \rightarrow - \omega_{\mu_1, \mu_2 \mu_3}
\ee
due to the signature difference between \cite{halfnull} and \cite{adsbh}.

In order to evaluate the Killing spinor equations acting on a generic spinor
({\ref{genspin}}) in the background of a 1/4-supersymmetric near-horizon geometry,
we define the following null basis:

\bea
{\bf{e}}^+ &=& -du
\nn
{\bf{e}}^- &=& dr +rh -{1 \over 2} r^2 \Delta^2 du
\nn
{\bf{e}}^1 &=& Z
\nn
{\bf{e}}^2 &=& {1 \over \sqrt{2}} (Z^2+i Z^3)
\nn
{\bf{e}}^\bt &=& {1 \over \sqrt{2}} (Z^2-i Z^3) \ .
\eea

It is then straightforward to evaluate all the components of the Killing spinor equation, making use
of the conditions on the near-horizon solutions found in \cite{adsbh} and summarized in Section 2.
We analyse the Killing spinor equation, first by integrating the $+$ and the $-$ components to solve for
the Killing spinor, and then by evaluating the integrability conditions associated with the remaining three components.

\subsection{Analysis of $-$ and $+$ Components}

We first analyse the $-$ component. This component implies that
\be
\partial_- \lp=0, \qquad \partial_- \lpb =0
\ee
and
\bea
\partial_- \lm + (-{1 \over \sqrt{2}} \Delta - {\sqrt{2}}i \ell^{-1}) \lp &=&0
\nn
\partial_- \lmb - {1 \over \sqrt{2}} \Delta \lpb &=&0 \ .
\eea
Hence $\lp$, $\lpb$ are independent of $r$, and
\bea
\lm &=& r  ({1 \over \sqrt{2}} \Delta + {\sqrt{2}}i \ell^{-1}) \lp + \mu^1_-
\nn
\lmb  &=& {r \Delta \over \sqrt{2}} \lpb + \mu^{\bar{1}}_-
\eea
where $\mu^1_-$, $\mu^{\bar{1}}_-$ are also independent of $r$.

Next, we analyse the $+$ component. On substituting the above conditions into this component, and
expanding out the resulting expressions in powers of $r$ we find:
\bea
- \partial_u \lp + (-{i \over \sqrt{2}} h_1 + {1 \over \sqrt{2}} \Delta -{\sqrt{2}}i \ell^{-1}) \mu^1_-
+ i h_2 \mu^{\bar{1}}_- &=&0
\nn
- \partial_u \lpb + i h_\bt \mu^1_- + ({i \over \sqrt{2}} h_1 + {1 \over \sqrt{2}} \Delta) \mu^{\bar{1}}_-
&=&0
\eea
together with the algebraic conditions
\be
(\ell^{-1} h_1 + {3 i \Delta \over \ell} + 2 \ell^{-2}) \lp =0,  \qquad h_\bt \lp =0 \ .
\ee
If $\lp \neq 0$, then the above algebraic conditions imply that $h_2 =0$, $\Delta=0$ and
$h_1 = -2 \ell^{-1}$. The solution is then simply $AdS_5$ with $F=0$.
As we are interested in solutions other than this, we shall henceforth
set $\lp =0$. Also note that this implies that $\lm$ is independent of $r$.

On substituting these conditions back into the $+$ component of the Killing spinor equations
we then find the conditions
\be
\partial_u \lm =0, \qquad \partial_u  \mu^{\bar{1}}_- =0
\ee
together with
\bea
(i \partial_1 \Delta - {i \over 2} \Delta h_1+{3i \Delta \over \ell} + {1 \over 2} \Delta^2)
\lm + (-\sqrt{2} i \partial_2 \Delta +{i \over \sqrt{2}} \Delta h_2 )\mu^{\bar{1}}_- &=&0
\nn
\sqrt{2} \partial_\bt \Delta \lm + \partial_1 \Delta \mu^{\bar{1}}_- &=&0 \ .
\eea

\subsection{Analysis of Remaining Components}

Combining all of the previous conditions, and integrating them up, we find
that either the solution is $AdS_5$, or
\bea
\lp &=& 0
\nn
\lpb &=& u \big( i h_\bt \lm + ({i \over \sqrt{2}} h_1 + {1 \over \sqrt{2}} \Delta) \mu \big) + \sigma
\nn
\lmb &=& {r \Delta \over \sqrt{2}}  \bigg(u \big( i h_\bt \lm + ({i \over \sqrt{2}} h_1 + {1 \over \sqrt{2}} \Delta) \mu \big) + \sigma \bigg) + \mu
\eea
where $\lm, \mu, \sigma$ are independent of $r$ and $u$.
We also find the following algebraic conditions on $\lm, \mu$:

\be
\label{alg1}
(-{i \over \sqrt{2}} h_1 +{1 \over \sqrt{2}} \Delta - \sqrt{2} i \ell^{-1}) \lm
+ i h_2 \mu =0
\ee
\be
\label{alg2}
(i \partial_1 \Delta - {i \over 2} \Delta h_1 + {3 i \Delta \over \ell} + {1 \over 2} \Delta^2) \lm
+(-\sqrt{2} i \partial_2 \Delta + {i \over \sqrt{2}} \Delta h_2) \mu =0
\ee
\be
\label{alg3}
\sqrt{2} \partial_\bt \Delta \lm + \partial_1 \Delta \mu =0 \ .
\ee
Note in particular, that if $\lm = \mu =0$ then the set of solutions in this
class can be at most $1/4$-supersymmetric. As we are interested in solutions preserving
exactly $1/2$ of the supersymmetry, we discard this case.

\subsubsection{Analysis of $1$ Component}

This component is equivalent to
\be
\label{aux1}
\partial_1 \lpb =0
\ee
and
\bea
\label{aux2}
\partial_1 \lm &=& \sqrt{2} h_2 \mu +({2 \sqrt{3} i \over \ell} a_1+ {1 \over \ell}) \lm
\nn
\partial_1 \mu &=& -\sqrt{2} h_\bt \lm \ .
\eea

On substituting ({\ref{aux2}}) into ({\ref{aux1}}), and making use of the previous
conditions, we find
\be
\partial_1 \sigma =0
\ee
together with
\be
\label{alg4}
\big( i \nabla_1 h_\bt + (-{1 \over 2} \Delta  +{i \over \ell}) h_\bt \big) \lm
+ \big( {i \over \sqrt{2}} \nabla_1 h_1 +{1 \over \sqrt{2}} \partial_1 \Delta \big) \mu =0
\ee
where here $\nabla$ denotes the Levi-Civita connection on the horizon.

\subsubsection{Analysis of $2$ and $\bt$ components}

These components are equivalent to:
\be
\label{aux3}
\partial_2 \lpb =0, \qquad \partial_\bt \lpb =0
\ee
and
\bea
\label{aux4}
\partial_2 \lm &=& (h_2 + {2 \sqrt{3} i \over \ell} a_2) \lm
\nn
\partial_2 \mu &=& (\sqrt{2} h_1 + 2 \sqrt{2} \ell^{-1}) \lm - h_2 \mu
\nn
\partial_\bt \lm &=& (-h_\bt +{2 \sqrt{3} i \over \ell} a_\bt) \lm -(\sqrt{2} h_1 +3 \sqrt{2} \ell^{-1}) \mu
\nn
\partial_\bt \mu &=& h_\bt \mu \ .
\eea
On substituting ({\ref{aux4}}) into ({\ref{aux3}}) and making use of the
previous conditions, we find
\be
\partial_2 \sigma =0, \qquad \partial_\bt \sigma =0
\ee
and
\bea
\label{alg5}
\big(i \nabla_2 h_\bt +{1 \over 2} \Delta h_1 -{i \over \ell}h_1 +2 \Delta \ell^{-1} \big) \lm
\nn
+ \big({i \over \sqrt{2}} \nabla_2 h_1 + h_2 (-{1 \over 2 \sqrt{2}} \Delta  +{3i \over \sqrt{2} \ell})
+{1 \over \sqrt{2}} \partial_2 \Delta \big) \mu =0
\eea
and
\be
\label{alg6}
i \nabla_\bt h_\bt \lm + \big( {i \over \sqrt{2}} \nabla_\bt h_1
+{1 \over \sqrt{2}} \partial_\bt \Delta + {1 \over 2 \sqrt{2}} \Delta h_\bt
-{3 \over \sqrt{2}} i \ell^{-1} h_\bt \big) \mu =0 \ .
\ee

We remark that it is straightforward to show that the solutions
preserve at least $1/4$ of the supersymmetry, because all of the conditions
are satisfied by setting $\lm = \mu =0$ and
taking $\sigma$ to be a non-zero complex constant.
For the simplest, supersymmetric $AdS_5$ regular black hole solution first constructed in \cite{adsbh}, it is also straightforward to see explicitly how the supersymmetry is enhanced from 1/4 to 1/2. An additional spinor
is obtained by setting $\lm = \sigma=0$ and taking $\mu$ to be a
complex constant, together with $\Delta$ constant and $h = -{3 \over \ell} Z$.

\appendix{The Integrability Conditions}

It is then straightforward to evaluate the integrability conditions associated with the
conditions ({\ref{aux2}}) and ({\ref{aux4}}). On combining these with the algebraic
conditions ({\ref{alg1}}), ({\ref{alg2}}), ({\ref{alg3}}), ({\ref{alg4}}), ({\ref{alg5}}), ({\ref{alg6}}),
and removing those conditions which are linearly dependent, one obtains the following conditions:

\be
\label{ex1}
(-{i \over \sqrt{2}} h_1 +{1 \over \sqrt{2}} \Delta - \sqrt{2} i \ell^{-1}) \lm
+ i h_2 \mu =0
\ee
\be
\label{ex2}
(\partial_1 \Delta +4 \ell^{-1} \Delta)\lm -\sqrt{2} \partial_2 \Delta \mu =0
\ee
\be
\label{ex3}
\sqrt{2} \partial_\bt \Delta \lm + \partial_1 \Delta \mu =0
\ee
\be
\label{ex4}
(i \psi_{1 \bt} +i \ell^{-1} h_\bt) \lm +({i \over \sqrt{2}} \psi_{11} +{1 \over 2 \sqrt{2}} \partial_1 \Delta) \mu =0
\ee
\be
\label{ex5}
(i \psi_{2 \bt} +i \ell^{-1} h_1 + {1 \over 4} \partial_1 \Delta +4i \ell^{-2}) \lm
+({i \over \sqrt{2}} \psi_{12} - {i \over \sqrt{2}} \ell^{-1} h_2) \mu =0
\ee
\be
\label{ex6}
i \psi_{\bt \bt} \lm +({i \over \sqrt{2}} \psi_{1 \bt} +{1 \over 2 \sqrt{2}} \partial_\bt \Delta
-{3 \over \sqrt{2}} i \ell^{-1} h_\bt) \mu =0
\ee
\be
\label{ex7}
(\psi_{12}+{i \over 2} \partial_2 \Delta + \ell^{-1} h_2) \lm - \sqrt{2} \psi_{22} \mu =0
\ee

where
\be
\psi_{ij} = \nabla_{(i} h_{j)} \ .
\ee

\appendix{Chong et. al Solution}

The near-horizon geometry of the cohomogenity-2 BPS black holes of Chong et al. \cite{chong1, chong2} has near-horizon data~\cite{KLR}:
\begin{eqnarray}
\label{NHmetricA}
 ds^2_{{\cal{S}}} &=& \frac{\ell^2 \Gamma d\Gamma^2}{4 P(\Gamma)} + \left( C^2 \Gamma - \frac{\Delta_0^2}{\Gamma^2} \right) \left( dx^1 + \frac{\Delta_0  (\alpha_0 - \Gamma)}{C^2 \Gamma^3 - \Delta_0^2} dx^2 \right)^2 + \frac{4 \Gamma P(\Gamma)}{\ell^2 (C^2 \Gamma^3 - \Delta_0^2)} (dx^2)^2 \ , \nn
\end{eqnarray}
where
\begin{equation}
P(\Gamma) = \Gamma^3 - \frac{C^2 \ell^2}{4} \left( \Gamma- \alpha_0 \right)^2 - \frac{\Delta_0^2}{C^2}
\end{equation}
with $C$ and $\Delta_0$ positive constants and $\alpha_0$ an arbitrary constant.
Furthermore,
\be
 \Delta &=& \frac{\Delta_0}{\Gamma^2}
 \ee
 and
\be
 h = \Gamma^{-1} \bigg( \left( C^2 \Gamma - \frac{\Delta_0^2}{\Gamma^2} \right)
  \left( dx^1 + \frac{\Delta_0  (\alpha_0 - \Gamma)}{C^2 \Gamma^3 - \Delta_0^2} dx^2 \right)  - d\Gamma \bigg)
 \ee
and
\begin{equation}
Z = \frac{\ell(\alpha_0 - \Gamma)C^2}{2\Gamma} dx^{1} + \frac{2\Delta_{0}}{\ell C^2\Gamma} dx^{2} +\frac{\ell}{2 \Gamma} d\Gamma \ .
\end{equation}


\end{document}